\documentclass{article}
\usepackage{authblk}
\usepackage{amsthm}
\usepackage{amssymb}
\usepackage{amsmath}
\usepackage{amsfonts}
\usepackage{wrapfig}
\usepackage{pgfpages}
\theoremstyle{plain}
\newtheorem{theorem}{Theorem}
\newtheorem{conclusion}{Conclusion}

\theoremstyle{definition}
\newtheorem{definition}[theorem]{Definition}

\newtheorem{question}[theorem]{Question}

\newcommand{\R}{\mathbb{R}}

\begin{document}

\title{Nonchaotic Models and Predictability of the Users' Volume Dynamics on Internet Platforms}

\author{Victoria Rayskin\thanks{
{I am very grateful to Professor M.W. Hirsch for important information, which helped to improve this paper.}
\\
{Research was sponsored by the Army Research Office and was
accomplished under Grant Number W911NF-19-1-0399. The views and conclusions contained in this
document are those of the authors and should not be interpreted as representing the official policies, either
expressed or implied, of the Army Research Office or the U.S. Government. The U.S. Government is
authorized to reproduce and distribute reprints for Government purposes notwithstanding any copyright
notation herein.}
\\
{I am very grateful tor the inspiring and stimulating work environment of the AMiTaNS conference, where some results of this paper were discussed and contributed to the Conference Proceedings.}}
\\
  Tufts University, Department of Mathematics,
Bromfield-Pearson Hall,
503 Boston Avenue,
Medford MA 02155, USA.\\
victoria.rayskin@tufts.edu
}

\date{\today}

\maketitle
\begin{abstract}
Internet platforms' traffic defines important characteristics of platforms, such as price of services, advertisements, speed of operations. The traffic is usually estimated with the help of the traditional time series models (ARIMA, Holt-Winters, etc.), which are successful in short term extrapolations of sufficiently denoised signals.   

We propose a dynamical system approach for the modeling of the underlying process. The method allows to discuss the global qualitative properties of the dynamics' phase portrait and long term tendencies. The proposed models are nonchaotic, the long term prediction is reliable, and it explains the fundamental properties and trend of various types of digital platforms. Because of these properties, we call the flow of these models the {\it trending flow}.  Utilizing the new approach, we construct the two-sided platform models for the volume of users, that can be applied to Amazon.com, Homes.mil or Wikipedia.org. 

We consider a generalization of the two-sided platforms' models to multi-sided platforms. If the equations' are cooperative, the flow is trending, and it helps to understand the properties of the platforms and reliably predicts the long term behavior.

We show how to reconstruct the governing differential equations from time series data. The external effects are modeled as system's parameters (initial conditions). 
\end{abstract}

\section{Introduction}\label{section-intro}

Internet platforms become one of the most popular objects for our daily activities.
There are many examples of interactions through platforms: health exchange platforms connect health insurance providers \& subscribers, bidding platforms connect auction buyers \& sellers, real estate platforms connect renters \& home owners, the `1000 genomes' platform connects multi-disciplinary researchers. The first three of these are examples of two-sided platforms and the last example is the example of multi-sided platform (\cite{RT}, \cite{EPA}). 

Users' traffic is one of the most important characteristics of platforms. For example, it helps the platform owners to price the platform's services, to predict platform's crashes due to high traffic and to plan platform development. It becomes increasingly important to model the dynamics of the platform's volume of users.

As discussed in~\cite{GJB}, usually, model's selection depends on complexity of the data, which spans a wide range. Easily predictable time series can be, for example, series with some seasonality, ``trend'' or their combination. On the opposite end of this range are series, generated by the process which transmit no information from the past to the future, like white noise processes. Series in the mid-range, generated by some non-linear process, pose interesting modeling challenges. It is important to quantify where on the complexity range a given time series falls. The detailed discussion of the complexity of the data, estimated with entropy, and model's selection can be found in~\cite{GJB}. Most of the real-life processes are non-linear and can be even ``chaotic''. As argued in~\cite{BK}, non-linear time series analysis is important and should be reflected in the model's selection.

We discuss the dynamical systems' approach for the study of the volume of users interacting through digital platforms. These models are generally non-linear (except for the Example~\eqref{linear-equations-system}), they have sensitivity to initial conditions, but they are not chaotic (we adopt R. Devaney~\cite{D} definition of chaos). 
The flow of the models discussed in this paper has such geometric properties, that make their long term behavior predictable. Some of the models are monotone (or even cooperative). The discussion of monotone and cooperative systems, and its comparison to chaotic dynamics can be found in~\cite{H1}, \cite{H5}, \cite{HS}, \cite{S} and in many references therein. The monotonicity of the models helps to understand global properties of the flow. Even though not all of the systems discussed in this work are monotone, in the domain of our interest, they have only trivial (period 1) periodic cycles (which are not dense), i.e., the systems are nonchaotic. The long term prediction of these systems is reliable and shows the trend of the flow. For this reason, we call the flow of our models the {\it trending} flow.

Our goal is to understand the platform-specific rules that govern the platform dynamics. We would like to understand the platforms' trends and long term tendencies.
The dynamical systems approach for platform analysis discussed in \cite{R1} and \cite{R2} makes it now possible to develop strategies for increasing platforms' efficiency inn the long run.

Furthermore,  in the study of multi-sided platforms (which becomes increasingly important), dynamical systems are very well suited for a high number of platform sides.

Finally, prediction of a platform's future behavior may be sensitive to noise, external effects and platform's policies. Any regime shift may significantly effect a prediction. ``Detecting regime shifts -- and adapting prediction models accordingly -- is an important area of future work'' (\cite{GJB}). The dynamical systems approach allows one to associate changes in the external world with the changes in the trajectories' initial conditions. Also, one can study regions of stability and other fundamental  qualitative concepts less sensitive to the external effects.

Behavior of non-linear systems can be chaotic and the long term prediction of trajectories often is not reliable. For this reason, we propose to fit the data into the models, which have trending flow, defined in Section~\ref{section-trending-flow}. This allows us to perform qualitative analysis of the processes and to understand the tendency of the processes (see Sections~\ref{section-approximation-examples} and \ref{section-Internet}). 
 Moreover, we can switch between different trajectories of the dynamical system to reflect the changes in the external conditions. We can also associate these switches with various incentives that help to influence platform's efficiency.

In Section~\ref{section-statistics} we discuss a method for deriving the vector field (of the specified form), which fits the time series data of the process.

In  Section~\ref{section-approximation-examples} we provide examples of the reconstructed dynamical systems and perform qualitative analysis of the phase portraits.

In Section~\ref{section-Internet} we discuss specific Internet platform models and the characteristics of the mechanism that governs the dynamics of these models. In particular, we compare the properties of the ``seller-buyer'' type of platforms and platforms like Wikipedia. We note that the competition between ``sellers'' as well as competition between ``buyers'' restricts growth of the number of users of a "seller-buyer" type of platform; while Wikipedia's popularity can grow (if there are no "wars" between contributors). 

Such analysis of the global properties of the platforms is possible because of the nonchaotic (trending) nature of the models' phase portraits. 

\section{Dynamics with trending flow}\label{section-trending-flow}
Many laws of nature and human activities can be modeled with the help of differential equations. Frequently, as illustrated with the next example, it is easier to observe and discover  these equations than their solutions.

When studying the physical systems, meteorological dynamics or chemical reactions that can be described with the help of the Lorentz equations, we face the complexity of chaotic behavior. The differential equations of this process are rather simple,  quadratic, and easily discoverable with the modern techniques:
\begin{equation}\label{eqn-lorenz}
\left\{
\begin{array}{l}
x_1'=10(x_2-x_1),\\
x_2'=x_1(28-x_3)-x_2,\\
x_3'=x_1x_3-\frac{8}{3} x_3
\end{array}
\right.
\end{equation}
 However, if we try to describe the Lorentz attractor trajectory, corresponding to the solution of these equations, we would not be able to write an explicit equation and would not be able to achieve high accuracy with approximations.

Many processes, for which the state variables' equations cannot be explicitly written due to their complexity, can be approximated with simple differential equations, which have very few non-linear terms.

Our goal is to study the complex processes with the help of differential equations that can be fitted into the data, generated by the processes.

This approach allows to understand the general laws governing the process, to perform qualitative analysis of the system's (global) phase portrait and also, depending on the external state, to choose one of the initial value's trajectory for the short term prediction of the process.

The term ``chaos'' is not always understood in the same sense. We will adopt the definition of R. Devaney \cite{D}, which requires, as one of the conditions of chaos,  density of periodic points. In that sense, many systems with ``small'' non-linearity are chaotic (such as system~\eqref{eqn-lorenz}), but the non-linear models that we study in this paper are nonchaotic (they only have stationary points, which are not dense in the state space). More over, some of these dynamical systems are monotone. The behavior of the monotone systems has been studied in multiple articles. In many cases it can be proved that their typical trajectories tend towards fixed points or periodic orbits, making the long term prediction more reliable. For the overview and comparison of chaotic and monotone dynamics see~\cite{H1}, and for greater details see \cite{H5}, \cite{HS}, \cite{S} and many references therein. The models discussed in this paper are not necessarily monotone, but they have very simple  geometry. 
\begin{definition}
Assume that we are interested in the dynamics on the set $D$, which is the closer of an open set. Consider a system of differential equations defined on the domain $D$ (possibly, well-defined only on the interior side of the boundary). 
We will say that the system of differential equations has {\it trending flow on $D$} if its semiflow ($t\geq 0$) with any initial value in $D$ either converges to a fixed point in $D$, or escapes the domain $D$ (in finite time). 
\end{definition}
Through this paper, the domain of interest is $D=[0,1]^n \subset \R^n$. The non-linear dynamical system models discussed in this paper 
have trending flow on the $D$. The study of the phase portrait of this flow can help to understand the long term forecast and trends of the processes. The forecast is reliable, because the flow is not chaotic on $D$.

\section{Techniques of reconstruction of the vector field from the data}\label{section-statistics}
How to reconstruct dynamical systems from data? We will follow the ideas of the SINDy method, developed by Brunton, Proctor, Kutz (\cite{BPK}). However, instead of trying to reconstruct one single trajectory, we will assume that the data comes from many different trajectories (because the process is influenced by external effects and consequently jumps from one initial value solution to another one). We will try to reconstruct the vector field for some region of these trajectories.

Accordingly, we will use LASSO, Ridge Regression, or a similar sparse regression for the vector field approximation, using the data points $\bar{x}(t_j)$. Here $t_j=1,...,p$ are the moments of times, when the state variables $\bar{x}(t_j)$ were recorded.

We will approximate the vector field $\bar{x'}(t_j):=\left(\bar{x}(t_j)-\bar{x}(t_{j-1})\right)/\left( t_j-t_{j-1}\right)$ by a linear combination of functions from the library 
\begin{equation*}\label{eqn-library}
{\bf L(X})=\{L_k(\bar{x}(t_j)) \}_{k,j=1}^{K,p}.
\end{equation*} 
(Here, $K$ is the number of functions included in the library ${\bf L}$.)

More specifically, for an $N$-dimensional system, we want to find the optimal values $A=\{\hat{a}_{ki}\}_{k,i=1}^{K,N}$ s.th. ${\bf L(X}) \cdot \widehat{A}$ approximates $({\bf X'})^T$:
\begin{equation*}\label{eqn-stat-approximation}
x'_i(t_j) \approx \widehat{x'_i(t_j)}=\sum_k L_k(\bar{x}(t_j))\cdot \hat{a}_{ki}.
\end{equation*}

The library ${\bf L}$ of "candidate" functions  may consist of homogeneous monomials, trigonometric and/or other simple functions. The matrix $\widehat{A}$ is sparse, because we need only a few terms for the vector field approximation. 

\section{Examples of the vector field reconstruction}\label{section-approximation-examples}
In this section, we consider several examples of differential equations reconstructed (precisely or approximately) from data, using the method, discussed in Section~\ref{section-statistics}.  These examples are related to the dynamics of the volume of users on Internet platforms, which we will present in Section~\ref{section-Internet}.

Our first example will be based on the data, simulated by the monotone system of differential equations
\begin{equation}\label{sqrt-equations-system}
\left\{
\begin{array}{l}
x' = -x +\sqrt{y},\\
y' = -y +\sqrt{x}.
\end{array}
\right. 
\end{equation}

\begin{figure}
\centering
\includegraphics[scale=0.4]{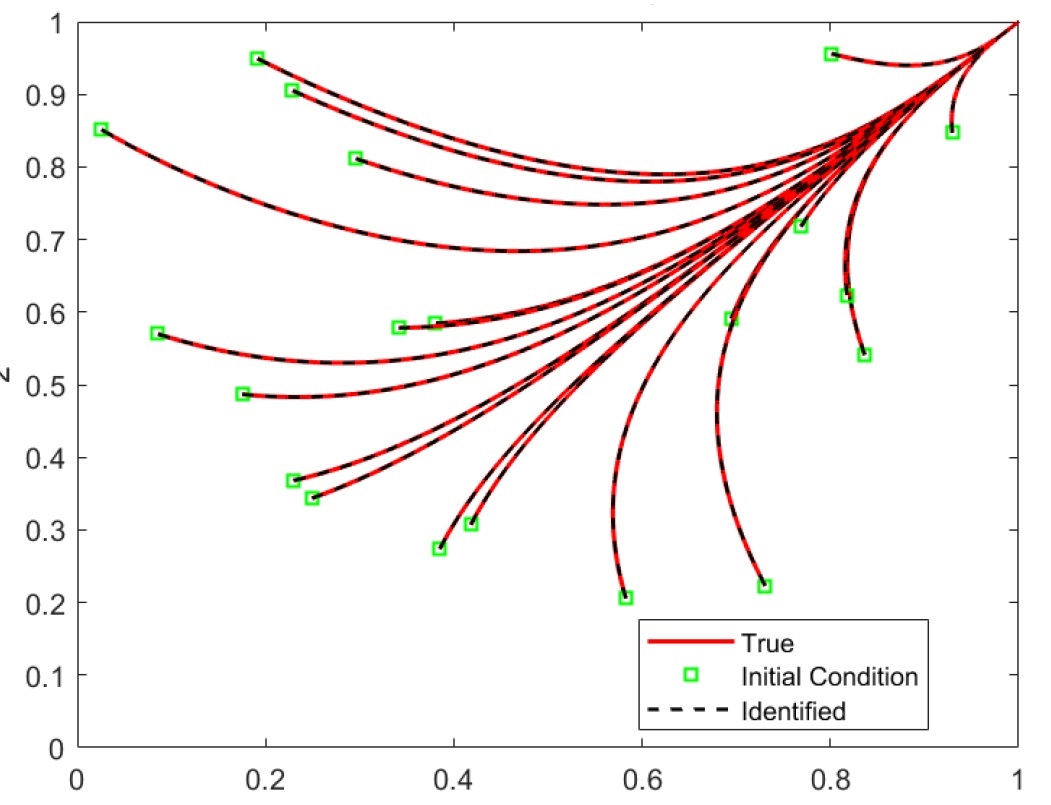}
\caption{True and identified trajectories, defined by the Equations~\eqref{sqrt-equations-system}. All trajectories in the basin of attraction $[0,1]^2\setminus {0}$ converge to $(1,1)$. The flow is trending. The trajectories' initial conditions are shown as green squares. The figure is reproduced with permission from AMiTaNS Conference Proceedings.\label{pic-flow-sqrt}}
\end{figure}

For the library ${\bf L}$, we will use monomials of order $\leq 5$ and $\sqrt{x}, \sqrt{y}$. The LASSO regression is capable of reconstructing the Equations~\eqref{sqrt-equations-system} precisely. The reconstructed dynamical system
can be analyzed, the global properties can be described, and the short term predictions can be performed starting at appropriate initial condition. The phase portrait of this system, which has 1 basin of attraction with the attracting fixed point $(1,1)$ is shown in Figure~\ref{pic-flow-sqrt}. The phase portrait analysis provides a reliable information for the long term prediction, because the system has trending flow.

\begin{figure}
\centering
\includegraphics[scale=0.4]{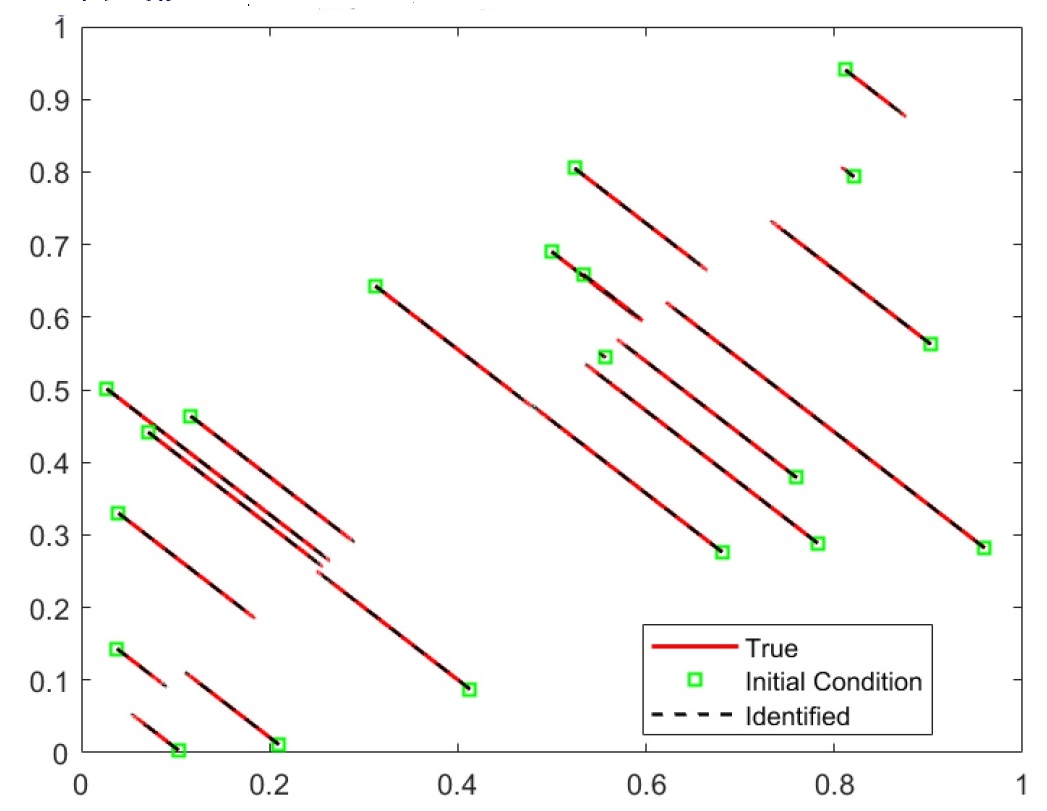}
\caption{True and identified trajectories, defined by the Equations~\eqref{linear-equations-system}. All trajectories converge to the fixed points on the diagonal $y=x$. The flow is trending. The trajectories' initial conditions are shown as green squares.  The figure is reproduced with permission from AMiTaNS Conference Proceedings.}
\label{pic-flow-linear}
\end{figure}

Next, we will simulate the data with the help of another monotone system, which is linear:
\begin{equation}\label{linear-equations-system}
\left\{
\begin{array}{l}
x' = -x +y,\\
y' = -y +x.
\end{array}
\right. 
\end{equation}
Theoretically, this system is very simple. However, it has the whole diagonal of stationary points. This singularity may cause problems with the traditional techniques for predicting future values of the process. Utilizing the dynamical systems approach, we can try to reconstruct the differential equations from the simulated data, using the same library ${\bf L}$ as in the Example~\ref{sqrt-equations-system}. Then, we can recover the equations precisely, and obtain the simple picture of the linear flow (Figure~\ref{pic-flow-linear}). And again, this analysis provides a reliable information for the long term prediction, because this system has trending flow.

Moreover, because we discovered that the process is linear, it is very easy to analyze this flow and to predict the future of the process.

In practice, the functions which we include in the library $L$ do not coincide with the functions (or other mechanisms) that generate the real process. In the next example, we will simulate the data with the help of the system, which consists of the functions not included in our library:
\begin{equation}\label{smoothed-step-system}
\left\{
\begin{array}{l}
x' = -x +y - \frac{1}{4\pi}\sin(4\pi y),\\
y' = -y +x -\frac{1}{4\pi}\sin(4\pi x)
\end{array}
\right. 
\end{equation}
We will use the library of monomials of order $\leq 5$, $\sin(nx)$ and $\cos(nx)$, $n=1,...,13$.

\begin{figure}
\centering
\includegraphics[scale=0.4]{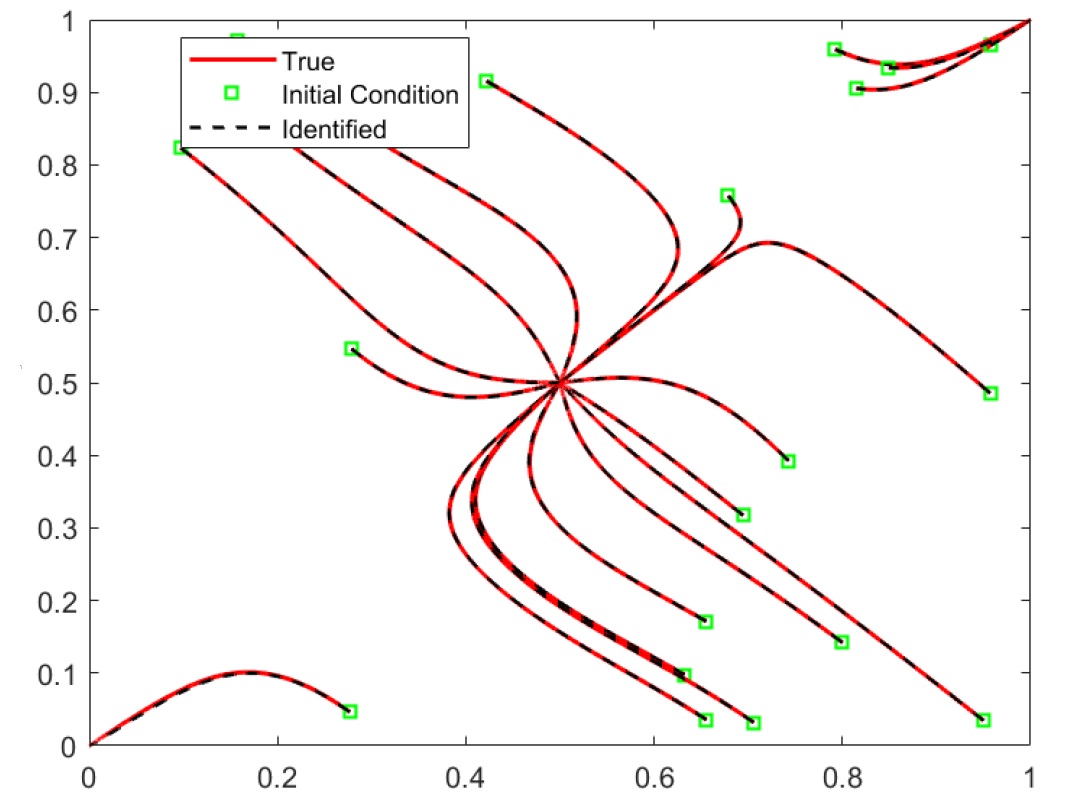}
\caption{Trajectories of the true Equations~\eqref{smoothed-step-system} and trajectories of the  approximation of these equations with the System~\eqref{smoothed-step-approxi-sytem}. There are 3 basins of attraction with the attracting stationary points $(0,0)$, $(.5,.5)$ and $(1,1)$. The flow is trending. The trajectories' initial conditions are shown as green squares.  The figure is reproduced with permission from AMiTaNS Conference Proceedings.}
\label{pic-flow-steps}
\end{figure}

The result of the LASSO regression is the system of the differential equations
\begin{equation}\label{smoothed-step-approxi-sytem}
\left\{
\begin{array}{l}
\begin{aligned}
x' =& -.2669x^5 +.2845y^5 -.3128\sin(3x)\\
     & +.2479\sin(3y) +.3808\cos(3x)\\
     & -.2753\cos(3y) -.0977\cos(5x)-...\\
y' =& -.2669y^5 +.6149\sin(3x)-.3128\sin(3y)\\
     & -.7437\cos(3x)+.3808\cos(3y)\\
     &+.9624\cos(5x) -.0977\cos(5y)-...
\end{aligned}
\end{array}
\right. 
\end{equation}
Even though the identified Equations~\eqref{smoothed-step-approxi-sytem} differ from the true Equations~\eqref{smoothed-step-system}, it appears that the approximation of the trajectories is of high precision (Figure~\ref{pic-flow-steps}).

In this example the true trajectories are generated by the monotone system (because $\frac{|\sin\alpha -\sin\beta|}{|\alpha -\beta|}\leq 1$), but we do not check monotonicity of the identified equations. Their phase portraits are similar. The phase portraits of such systems are described by the Theorem~\ref{thm-portrait}. Because the geometry of the phase portraits is simple, the long term tendency of the trajectories can be reliably estimated with the help of the phase portrait.


\section{Dynamics of the volume of users on Internet platforms}\label{section-Internet}
In this section, we will use the ideas of the differential equations approximation, discussed in the Sections~\ref{section-trending-flow},~\ref{section-statistics},~\ref{section-approximation-examples}, for the analysis of the traffic on Internet platforms. The estimates of the volume of platform users are important for the platform owners, helping them to price the platforms' services, to predict the moments of high traffic (which occasionally cause platform's crashes) and to understand the long term future of platforms' popularity.

We will focus on the examples of the dynamical system models for the two types of Internet platforms: those that involve interaction similar to the "seller-buyer" relations and those that are similar to the Wikipedia platform. We will analyze tendency and global properties of the volume of users interacting through these platforms with the help of the models, having trending flow. For more detailed analysis and proof of the results stated in this section see \cite{R1} and \cite{R2}.

First, we discuss the {\it two-sided} platforms. They are characterized by two different types of users: renters and home owners (on Homes.mil), sellers and buyers (on Amazon.com), readers and contributors (on Wikipedia.org), males and females (on Match.com), job seekers and employers (on Jobs.com), etc. 

The volume of users in our model is a fraction of all theoretically possible users on each side of the platform. We will denote them $b$  and $g$ (for example, $b$ may represent the fraction of all possible buyers and $g$ may represent the fraction of all possible sellers interacting through the Amazon.com platform) and consider the following system:
$$
\left\{
\begin{array}{l}
b' =V(g) + R(b),\\
g' =W(b) + S(g).
\end{array}\ \ \  b\in [0,1],\ g\in [0,1],
\right.
$$
Here, $R(b)$, $S(g)$ are the same-side network effects, and $V(g)$, $W(b)$ are the cross-side network effects. 

The same side network effect represents the interaction within the same type of users. This effect can be positive ($S:[0,1]\to [0,\infty)$), negative ($S:[0,1]\to (-\infty,0]$, or non-existent ($S\equiv 0$). The same for the $R$ function.

The negative effect is typical for renters and buyers who want to keep costs low, for home owners (on Homes.mil) sellers (on Amazon.com), job seekers (on Jobs.com), males or females (on Match.com) who prefer lower competition. Positive effect occurs between reviewers on Yelp, or gamers who share access tools on gaming platforms. On a platform like Wikipedia, the same-side network effect does not exist (unless there are some "edit wars" between contributors). 

In Subsection~\ref{section-seller-buyer-platform}, we consider platforms with the negative same-side network effect, such as Homes.mil, Amazon.com, Match.com, Jobs.com. Platforms with the negative same-side network effect we call "seller-buyer" type of platform.
 
Then, in Subsection~\ref{section-wiki-platform}, we discuss the Wikipedia platform. 

In all examples discussed below, we approximate the same-side network effects with linear functions. 

The cross-side network effect represents user’s preferences and interest in the other type of users. We believe that it reflects important differences between various platforms. Usually this effect is non-negative. Therefore, in this paper we assume $V\geq 0$ and $W\geq 0$. Also,  nobody wants to join a platform without presence of the opposite party. This implies the assumption $W(0)=V(0)=0$. Time and range re-scaling allow us to assume that $W,V : [0,1] \rightarrow [0,1].$ We will call $W$ and $V$ the attachment functions.

\subsection{Systems with negative same-side network effect}\label{section-seller-buyer-platform}
In this subsection, we discuss the users' traffic model for platforms with the negative same-side network effect, approximated by linear functions $-\epsilon b$ and $-\delta g$:
\begin{equation}\label{neg-same-side-system}
\left\{
\begin{array}{l}
b' = V(g) - \epsilon b,\\
g' =W(b) - \delta g,
\end{array}
\right.
\end{equation}
where 
$$b\in [0,1],\ g\in [0,1],\ \epsilon, \delta \geq 1 \mbox{  and  }$$
$$V, W: [0,1]\to [0,1],\ \  V(0)=W(0)=0.$$

It is easy to see that each trajectory defined by the Equations~\eqref{neg-same-side-system}, with its initial condition in the unit square $[0,1]^2$, remains in the square and converges to a fixed point (due to negative divergence there are no periodic orbits). In other words, the model is well defined (platform's population does not grow beyond $100\%$) \& the flow is trending. We will discuss the tendency (convergence to the fixed points) of the trajectories and other qualitative properties of the phase portrait of the system with various attachment functions $V$ and $W$.

The first example is the power attachment, i.e., 
\begin{equation}\label{eqn-power}
V(g)=g^{\alpha}, \ \ W(b)=b^{\alpha},\ \ \alpha \leq1.
\end{equation}

\begin{figure}
\centering
 \includegraphics[scale=0.3]{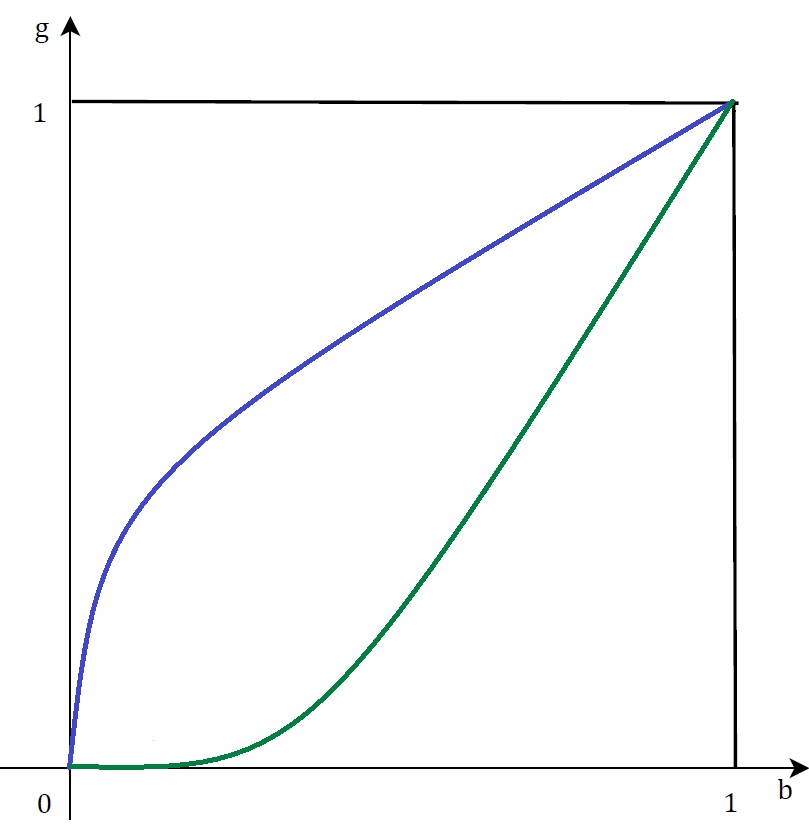}
\caption{The power attachment functions. The fixed points of the trending flow defined by the Equations~\eqref{neg-same-side-system} with the attachments expressed by the Equations~\eqref{eqn-power} are located at the intersections of the attachment functions: at the origin and at the corner $(1,1)$. This strong attachment attracts all theoretically possible users. Eventually, everybody joins the platform. See Figure~\ref{pic-flow-sqrt} of Section~\ref{section-approximation-examples}.  The figure is reproduced with permission from AMiTaNS Conference Proceedings.}
\label{fig-power}
\end{figure}

Because fixed points of the dynamical system~\eqref{neg-same-side-system} are located at the intersections of $V$ and $W$, we draw them on the same graph (Figure~\ref{fig-power}). The phase portrait is shown in the Figure~\ref{pic-flow-sqrt} of Section~\ref{section-approximation-examples}. It has one attracting fixed point at $(1,1)$. The quadrant $[0,1]^2\setminus \{0\}$ is the basin of attraction. This dynamics is monotone and, moreover, it has only simple fixed point attractor. The traffic on the corresponding platforms can be predicted relatively easy. We can conclude that with the power (very strong) cross-side attachment, all population tends to join the platform eventually.

Many platforms can be characterized by step-function attachments. Here, for simplicity, we show only 3 steps:
\begin{equation*}V(g)=\left\{
\begin{array}{ll}
0, & g\in[0,.25)\\
1/2, & g\in[.25,.75)\\
1, & g\in[.75, 1]
\end{array}
\right.
\end{equation*}
\begin{equation*}W(b)=\left\{
\begin{array}{ll}
0, & b\in[0,.25)\\
1/2, & b\in[.25,.75)\\
1, & b\in[.75, 1]
\end{array}
\right.
\end{equation*}
\begin{figure}
\centering
\includegraphics[scale=0.3]{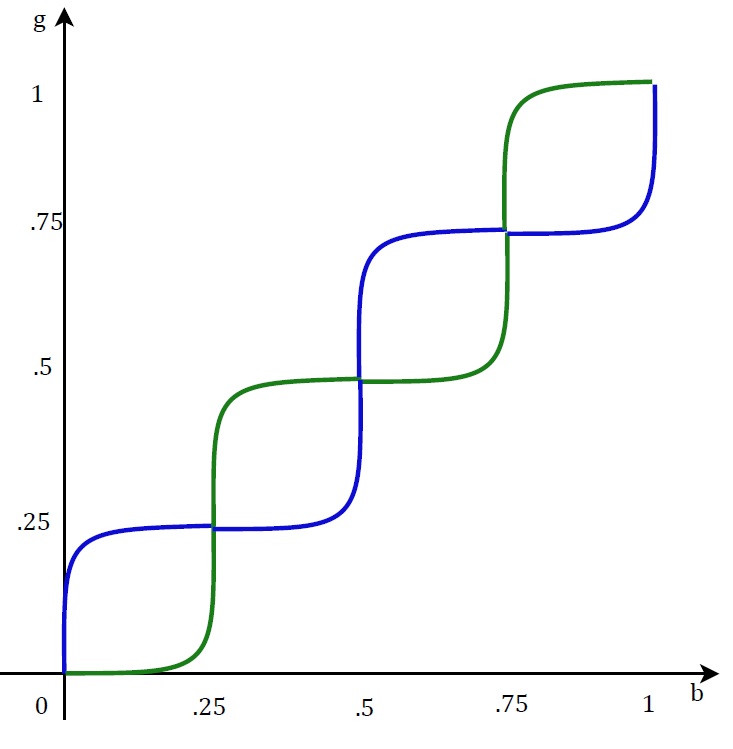}
\caption{Smoothed step attachment (with 3 steps) creates 3 attracting fixed points in the phase portrait of the dynamical system. They are located in the 1st, 3rd and 5th points of intercept of the attachments. The other 2 points of intercept are the hyperbolic fixed points of the dynamical system.  The figure is reproduced with permission from AMiTaNS Conference Proceedings.}
\label{fixed_pts_smoothed_step_fn}
\end{figure}
This dynamics is monotone, the attractors are simple fixed points (the flow is trending) and the traffic of the corresponding platforms can be reliably predicted. However, the attachment functions are discontinuous. For the real life applications, it is natural to assume that these steps are smoothly connected (possibly via rapidly growing functions called here $c_1$ and $c_2$). Thus, we will consider smoothed step-function attachments shown in Figure~\ref{fixed_pts_smoothed_step_fn}: 

\begin{equation}\label{eqn-smoothed-step}
V(g)=\left\{
\begin{array}{ll}
0, & g\in[0,.25-\delta]\\
c_1(g), & g\in (.25-\delta, .25 +\delta)\\
1/2, & g\in[\delta + .25,.75-\delta]\\
c_2(g), & g\in (.75-\delta, .75 +\delta)\\
1, & g\in[\delta +.75, 1]
\end{array}
\right.
\end{equation}
A similar equation can be written  for $W(b)$.

\begin{figure}
\centering
 \includegraphics[scale=0.4]{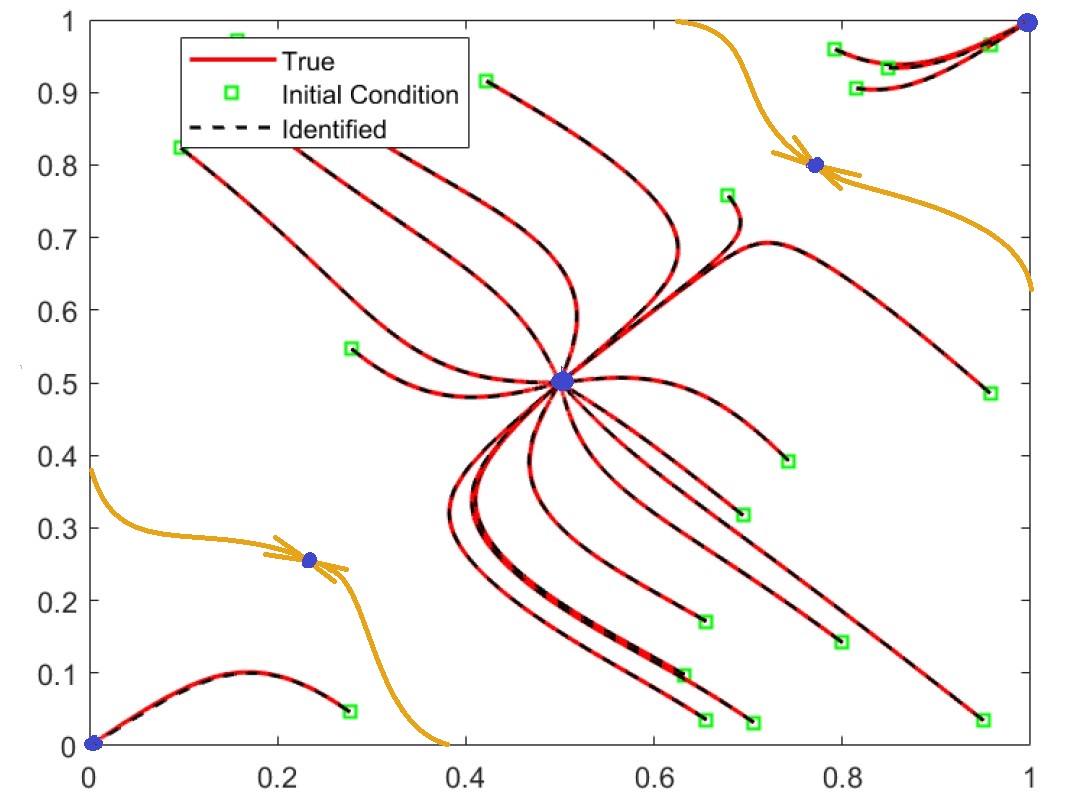} 
\caption{Fixed points are denoted by blue dots, initial values of trajectories are green. Orange separatrix, passing through saddle fixed points, separate 3 basins of attraction. Low volume of users (in the low-left basin) eventually extinct. If the volume of users is in the middle basin of attraction, the tendency of the volume is to the middle fixed point. 
If the level of users is in the upper-right basin of attraction, the platform will eventually attract all population. These global characteristics are reliable, because the flow is trending.}
\label{smooth_stairs_basins}
\end{figure}

The phase portrait, corresponding to the smoothed step attachment (with 3 steps) is illustrated by the Figure~\ref{pic-flow-steps} of the Section~\ref{section-approximation-examples}. It has 3 basins of attraction containing stable fixed points. The basins are separated by separatrix passing through the 2 saddle fixed points. See Figure~\ref{smooth_stairs_basins}. Because we used the equations with the smoothed step attachment functions for the approximation of the equations with the step function attachment, it is natural to assume that $V$ and $W$ are non-decreasing functions. This makes the dynamics cooperative and easily predictable. The class of cooperative systems (see~\cite{HS}) has a variety of applications in biology, chemistry, physics and economics. However, even without monotonicity assumption, due to the negative same-side network effect, our model generates easily predictable behavior of the trajectories. If we consider a more general case of non-monotone attachments $V$ and $W$, Theorem~\ref{thm-portrait} implies that the flow of such system is trending.  
\begin{conclusion}\label{conclusion-sell-buy}
The model defined by the Equations~\eqref{neg-same-side-system} with attachments expressed by the Equations~\eqref{eqn-smoothed-step} has the trending flow shown in Figure~\ref{smooth_stairs_basins}. It has the following properties.
\begin{itemize}
\item If the volume of users is low (within the low-left basin), all existing users tend to leave the platform eventually. 
\item If the volume of users is in the middle basin of attraction, the tendency of the volume is the middle fixed point. 
\item If the level of users is high (in the upper-right basin), the platform will eventually attract all population. 
\item The long term prediction of the flow is reliable, because the flow is trending.
\item The jumps between the basins of attraction happen with the help of external effects. For example, the platform owner can offer some incentives or change the platform's policy to jump from the lower basin to the higher one. 
\end{itemize}
\end{conclusion}

The general dynamics, defined by the Equations~\eqref{neg-same-side-system},  can be described by the following theorem.
\begin{theorem}\label{thm-portrait}
Suppose the two-sided platform model described by the Equations~\eqref{neg-same-side-system} is smooth and has finite number of fixed points located at $\{(b,g)\in [0,1]^2: \epsilon b=V(g), \delta g= W(b)\}.$ Then, the fixed points are: stable nodes/saddles/saddle-nodes, and stable spirals (no repelling fixed points, i.e., at least one eigenvalue has negative real part).
\begin{itemize}
\item $(b_0,g_0)$ is a stable node/spiral, if and only if $V'(g_0)\cdot W'(b_0) <\epsilon \delta$;
\item $(b_0,g_0)$ is a saddle, if and only if $V'(g_0)\cdot W'(b_0) >\epsilon \delta$;
\item $(b_0,g_0)$ belongs to center manifold, if and only if $V'(g_0)\cdot W'(b_0) =\epsilon \delta$. Moreover, if $V''(g_0)\cdot W'(b_0) + \frac{1}{\epsilon}( V'(g_0))^2\cdot W''(b_0) \neq 0,$ then one of the branches of center manifold converges to $(b_0, g_0)$, but the other one diverges from this fixed point. 
\end{itemize}
\end{theorem}
The proof of this result can be found in \cite{R2}.

This theorem shows that the phase portrait's analysis of the models described by the Equations~\eqref{neg-same-side-system} helps to approximate the long term behavior of the trajectories. The trajectories of the Equations~\eqref{neg-same-side-system} like trajectories of monotone systems are nonchaotic. They have simple attractors (fixed points) and their long term prediction is reliable. 

Multi-sided platforms are gaining popularity and become an important research subject. We will model them with the multi-dimensional differential equations. The multi-sided platforms with the negative same-side network effect can be described by the following system:
\begin{equation}\label{neg-same-side-system-multi-sided}
\left\{
\begin{array}{l}
b_1' = V_1(b_2, b_3, ..., b_n) - \epsilon_1 b_1,\\
b_2' = V_2(b_1,b_3,..., b_n) - \epsilon_2 b_2,\\
\vdots\\
b_{n-1}' =V_{n-1}(b_1, b_2,...,b_{n-2},b_n) - \epsilon_{n-1} b_{n-1},\\
b_n' =V_n(b_1,...,b_{n-1}) - \epsilon_n b_n,
\end{array}
\right.
\end{equation}
where 
$$b_k\in [0,1],\ \epsilon_k \geq 1 \mbox{  and  }V_k: [0,1]\to [0,1],\ \  V_k(0,...,0)=0$$
$$\mbox{for }k=1,...,n.$$
Let us also assume that there is a finite number of stationary points in this system.

Some of the properties of this system are similar to the properties of Equations~\eqref{neg-same-side-system}. For example, the flow does not leave the $n$-dimensional unit cube.  It is easy to see that at each fixed point the sum of the eigenvalues of the linearized system is $-\sum_{k=1}^n \epsilon_k<0$, i.e., at least one eigenvalue is negative.
If we assume that attachment functions of the Equations~\eqref{neg-same-side-system-multi-sided}  approximate some discrete step function attachments (i.e., $V_i$ are non-decreasing), the system is cooperative and its long term prognoses is reliable.
However, in the general case, absence of closed orbits in multidimensional cases cannot be proved with the same technique as for the Equations~\eqref{neg-same-side-system}. The behavior of the trajectories can be complex and  their long term prediction may be unreliable. For the modeling of multidimensional processes the following question is important.

\begin{question}
Are there conditions on the attachment functions $V_k$ that make the flow of the equations~\eqref{neg-same-side-system-multi-sided} trending, but not necessarily monotone?
\end{question}

\subsection{Systems without the same-side network effect}\label{section-wiki-platform}
The model for platforms without the same-side network effect can be written as the following system of equations:
\begin{equation}\label{eqn-wiki}
\left\{
\begin{array}{l}
b' = V(g),\\
g' =W(b),
\end{array}
\right.
\end{equation}
where 
$$V, W: [0,1]\to [0,1] \mbox{ and } V(0)=W(0)=0.$$
If we assume that $V$ and $W$ are differentiable, and because $\frac{\partial \left(V(g)/b\right)}{\partial g} = \frac{\partial \left(W(b)/g\right)}{\partial b} =0$ (when $b\neq0$, $g\neq 0$), we can see that the system is cooperative.

Also, without the differentiability assumption, we can show that the flow is trending. Indeed, since the attachment functions by definition are non-negative, the number of users never declines on such platforms. Also, because $b$ and $g$ never decrease, there are no cycles. Applying these ideas to the general tendency of the popularity of Wikipedia, we can conclude that in the long run, it should increase, if there is no the same-side competition. 

The trajectories of the Equations~\eqref{eqn-wiki} satisfy
\begin{equation}\label{eqn-wiki-solution}
\int V(g)d g-\int W(b) db = 0 .
\end{equation}
 
If $V$  and $W$ vanish not only at the origin, then non-zero stationary points are formed, and the flow of users converges to such point(s) from left and below, while staying within the point's basins of attraction; or the flow escapes the unit square. Thus, this flow is trending.

\begin{figure}
\centering
\includegraphics[scale=0.3]{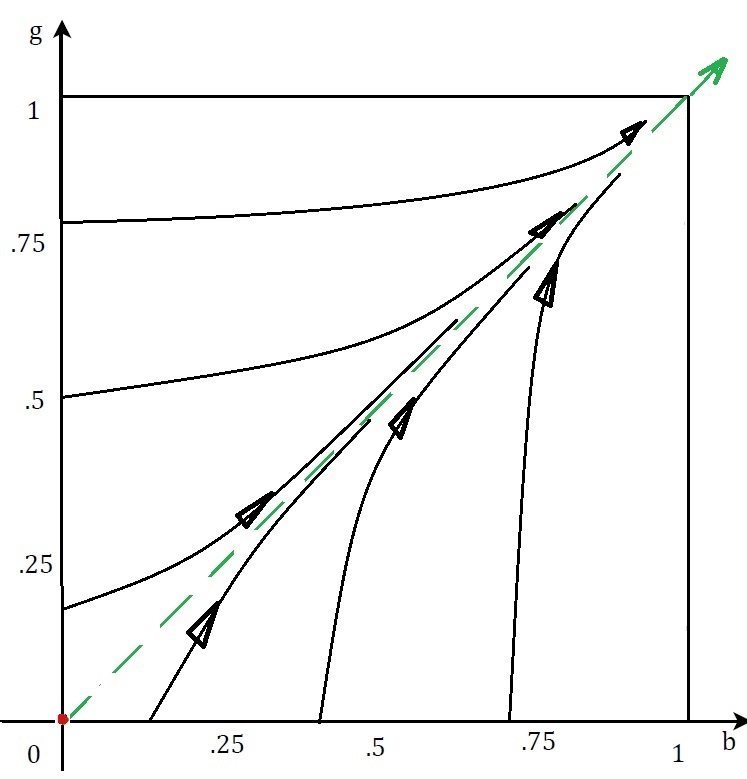} 
\caption{If we assume that negative same-side network effect is not present, the volume of platform users does not decline, unless it jumps to a lower trajectory (due to policy changes, for example). However, if the origin was the only fixed point (as shown here), the flow of users would escape the unit square.  The figure is reproduced with permission from AMiTaNS Conference Proceedings.}
\label{wiki-flow}
\end{figure}

If $0$ is the only stationary point, or the volume of users is either above the highest stationary point or to the right of the rightmost stationary point, the  trajectories would eventually escape the unit square $[0,1]^2$  as shown in Figure~\ref{wiki-flow}. Thus, we assume that either the negative interaction between the users of the same side must appear at some point and the model changes, or there are some non-zero stationary points (with both, $b$ and $g$ positive coordinates) and the initial volume of user is below and to the left of some stationary point.

The authors of \cite{KBMG},  \cite{HKKR},  \cite{BKM}, \cite{HGMR} discuss issues like Wikipedia's "edit wars" between contributors and newcomer retention. They also discuss some new policies, re-interpretations of the community rules and curating according to seniority.
This indicates some competition between contributors and possibly appearance of the  negative same-side network effect. In this case, the system becomes similar to the "seller-buyer" system discussed in the Section~\ref{section-seller-buyer-platform}. Then, incentives provided to the contributors (or some new policy) can move the dynamics into a higher basin of attraction, as discussed in Conclusion~\ref{conclusion-sell-buy}, and increase the volume of contributors and readers.

The recent analysis of \cite{HGMR} shows that the  number of Wikipedia contributors declines. If the "edit wars" are not fundamental characteristics of the Wikipedia platform, they can be viewed as a temporary external effects. In this case, the decline of the volume of users can possibly be explained as jumps from the higher trajectories to lower trajectory, while the fundamental law governing the dynamics has no negative same-side network effects and can be modeled with the Equations~(\ref{eqn-wiki}). Then, the long term tendency of the trajectories is the growth of the number of users.

Let us assume that the Equations~\eqref{eqn-wiki} describe the dynamics of the users' volume of  Wikipedia. As we noticed, in this case $V$ and $W$ must vanish not only at the origin, i.e., at some points $(g_i, b_j)\in (0,1]^2$ each side becomes indifferent to the opposite side:
\begin{equation*}
V(g_i)=0 \mbox{ and }W(b_j)=0.
\end{equation*}

\begin{figure}
\centering
\includegraphics[scale=0.4]{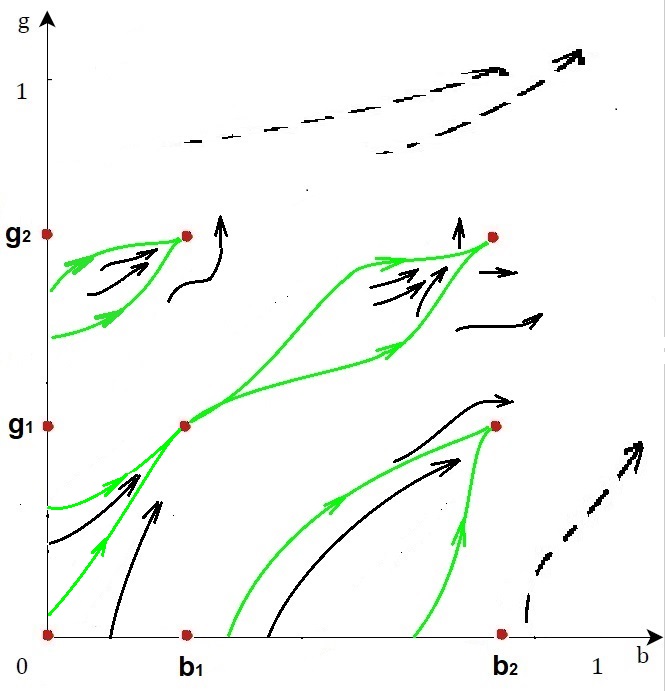}
\caption{If the attachment function $V(g)$ has three zeros (at $0$, $g_1$ and $g_2$) and the attachment function $W(b)$ has 3 zeros (at $0$, $b_1$ and $b_2$), the system has 9 stationary points shown in red color. The flow of users (with some positive volume of users on each side) increases approaching one of the 4 non-zero stationary points staying within some basin of attraction (shown in green), or escapes the unit square. If the flow was above either $b_2$ or $g_2$, (shown with dashed lines) it would escape the unit square. This flow is trending on a bounded domain.}
\label{fig-wiki-dynamics-with-indifference}
\end{figure}

Then, $(g_i, b_j)$ are non-zero stationary points. Since $V(g)\geq 0 $ on its domain, $g_i$ must be a point of local minimum of $V(g)$; and consequently $V'(g_i)=0$. Similarly, $W'(b_j)=0$. This implies that all non-zero stationary points $(g_i, b_j)$ belong to the  center manifolds, which direct the flow North-East, as shown in Figure~\ref{fig-wiki-dynamics-with-indifference}. Some of the trajectories may be trapped between the center manifolds and converge to the stationary points, while other trajectories escape the unit square. In the latter case, we assume that it takes very long time to escape the unit square (during a reasonably long time the volume of users does not grow beyond 100\%).

The Equations~\eqref{eqn-wiki} can be generalized for multi-sided platforms without same side network effect:
\begin{equation}\label{no-same-side-multi-sided}
\left\{
\begin{array}{l}
b_1' = V_1(b_2, b_3, ..., b_n) ,\\
b_2' = V_2(b_1,b_3,..., b_n) ,\\
\vdots\\
b_{n-1}' =V_{n-1}(b_1, b_2,...,b_{n-2},b_n) ,\\
b_n' =V_n(b_1,...,b_{n-1}) ,
\end{array}
\right.
\end{equation}
where 
$$b_k\in [0,1] \mbox{  and  }V_k: [0,1]\to [0,1],\ \  V_k(0)=0  \mbox{ for }k=1...n.$$

Because the attachment functions are non-negative, there are no cycles and the flow of users is always growing.
\begin{conclusion}\label{conclusion-wiki}
The platforms, modeled with the help of Equations~\eqref{no-same-side-multi-sided}(in particular, Equations~\eqref{eqn-wiki}) have trending flow with the following properties.
\begin{itemize}
\item In the long run, the volume of users does not decrease.
\item The long term prediction is reliable.
\item A temporary decline of the volume of users may be due to external effects. This corresponds to the jumps from the higher trajectories to the lower trajectories of the governing equations. 
\end{itemize}
\end{conclusion}


\begin{thebibliography}{KH}

\bibitem{BKM} I. Beschastnikh, T. Kriplean, D.W. McDonald {\it Wikipedian Self-Governance in Action: Motivating the Policy Lens}, ICWSM, 2008

\bibitem{BK} E. Bradley, H. Kantz, {\it Nonlinear time-series analysis revisited} (2015), Cahos,200

\bibitem{BPK} S.L. Brunton, J.L. Proctorb, J.N. Kutzc, {\it Discovering governing equations from data by sparse identification of nonlinear dynamical systems,} {Proceedings of the National Academy of Sciences,} {\bf 113}, (2016), 3932--3937.    

\bibitem{CM} Cowpertwait, P. S. P., \& Metcalfe, A. V. (2009) Introductory time series with R. New York: Springer-Verlag

\bibitem{D} R. Devaney, ``An introduction to chaotic dynamical systems,''  Benjamin/Cummings, Menlo Park, CA, (1986)

\bibitem{EPA} T.R. Eisenmann, G. Parker, M. van Alstyne, {\it Strategies for Two-Sided Markets,} Harvard Business Review, {\bf 84(10)}, (2006).

\bibitem{GJB} J. Garland, R. James,  E. Bradley, {\it Model-free quantification of time-series predictability} (2014) Phys. Rev. E {\bf 90(5)}. DOI: 10.1103/PhysRevE.90.052910

\bibitem{HKKR} A. Halfaker, A. Kittur, R. Kraut, J. Riedl, {\it A Jury of Your Peers: Quality, Experience and Ownership in Wikipedia} (2009). WikiSym Article 15, 10 pages. DOI:10.1145/1641309.1641332

\bibitem{HGMR}  A. Halfaker, R.S. Gieger, J. Morgan, J. Riedl {\it The Rise and Decline of an Open Collaboration System: How Wikipedia's reaction to sudden popularity is causing its decline} (2013) American Behavioral Scientist, {\bf 57(5)} 664--688, DOI:10.1177/0002764212469365 

\bibitem{H1} M. Hirsch, {\it Attractors for discrete–time monotone dynamical systems in strongly ordered spaces} Geometry and Topology: Lecture Notes in Mathematics 1167, 141--153. J. Alexander, J.Harer, editors. Springer-Verlag, New York, 1985.





\bibitem{H5} M. Hirsch {\it On the nonchaotic nature of monotone dynamical systems}, European Journal of Pure and Applied Mathematics, {\bf 12}, No. 3 (2019), 680--688

\bibitem{HS} M. Hirsch \& H. Smith, Monotone Dynamical Systems, ``Handbook of Differential Equations,'' volume 2, chapter 4. A. Ca{n}ada, P. Dra{b}ek \& A. Fonda, editors. Elsevier North Holland 2005.

\bibitem{KBMG} T. Krieplean, I. Beschastnikh, D.W. McDonald, S.A. Golder {\it Community, Consensus, Coercion, Control: CS*W or How Policy Mediates Mass Population}, presentation at GROUP 2007.

\bibitem{R1}V. Rayskin,  {\it Dynamics of two-sided markets, Review of Marketing Science}, {\bf 14}, (2016), 1--19.

\bibitem{R2}V. Rayskin,  {\it Users' dynamics on digital platforms}, {\em Mathematics and Computers in Simulation}, {\bf 142}, (2017), 82--97

\bibitem{RT}J.-C. Rochet,  J. Tirole, {\it  Platform competition in two-sided markets,} {Journal of the European Economic Association}, {\bf 1(4)}, (2003), 990--1029.

\bibitem{S} H. Smith, Monotone dynamical systems: reflections on new advances \& applications, Disc. \& Cont. Dyn. Systems Series B {\bf 37} (2017), no. 1, 485--504.

\end{thebibliography}
\end{document}